\documentclass[article,12pt]{article}
\usepackage{authblk}
\usepackage{mathrsfs}

\makeatletter %
  \renewcommand{\@biblabel}[1]{#1.}
\makeatother

\usepackage{overpic}
\usepackage{color}

\usepackage{geometry}
\usepackage{amssymb}
\usepackage[superscript,biblabel]{cite}
\title{Convective self-aggregation as a cold pool driven critical phenomenon}

\author[1]{Jan O. Haerter}

\affil[1]{Niels Bohr Institute, University of Copenhagen, Blegdamsvej 17, 2100 Copenhagen, Denmark}

\begin{document}

\maketitle
\noindent
{\bf Author affiliation}: Blegdamsvej 17, 2100 Copenhagen, Denmark. \\
{\bf Corresponding author}: Jan O. Haerter; haerter@nbi.ku.dk\\


\small
\noindent
{\bf 
Convective self-aggregation is when thunderstorm clouds cluster over a constant temperature surface in radiative convective equilibrium. 
Self-aggregation was implicated in the Madden-Julian Oscillation and hurricanes. 
Yet, numerical simulations succeed or fail at producing self-aggregation, depending on modeling choices. 
Common explanations for self-aggregation invoke radiative effects, acting to concentrate moisture in a sub-domain. 
Interaction between cold pools, caused by rain evaporation, drives reorganization of boundary layer moisture and triggers new updrafts. 
We propose a simple model for aggregation by cold pool interaction, assuming a local number density \boldmath$\rho(\mathbf{r})$ of precipitation cells, and that interaction scales quadratically with $\rho(\mathbf{r})$. 
Our model mimics global energy constraints by limiting further cell production when many cells are present. 
The phase diagram shows a continuous phase transition between a continuum and an aggregated state. 
Strong cold pool-cold pool interaction gives a uniform convective phase, while weak interaction yields few and independent cells.
Segregation results for intermediate interaction strength.
} 

\section*{Introduction}
Drawing a link between cloud and precipitation processes, and statistical mechanics is intriguing:
similar to the characteristics of critical phenomena\cite{yeomans1992statistical}, cloud fields often show long-ranged correlations and near-fractal scaling \cite{Cahalan:1989,Lovejoy:2006}. 
The possibility that the cloud field might have scaling properties is appealing because this might imply universal behavior, where only few features of the small scale interactions are relevant in capturing emergent organization \cite{goldenfeld2018lectures}. 
Invoking non-equilibrium statistical physics, a relation to the sand pile model \cite{bak1987self} has been drawn, which coins the notion of {\it self organized criticality} (SOC).
Loosely speaking, precipitation could be ascribed characteristics of an avalanche, when moisture, brought into a domain of interest, is abruptly released after a critical mixing ratio is exceeded \cite{Peters:2002,Peters:2006}.

A particularly appealing candidate for critical behavior is the radiative convective equilibrium framework \cite{held1993radiative}. 
Radiative convective equilibrium requires overall constant and homogeneous boundary conditions, e.g. constant surface temperature and moisture as well as insolation, and that the energy fluxes entering and leaving the system be equal. 
Under such conditions, humidity perturbations have repeatedly been shown to grow over time, in a process termed convective self-aggregation \cite{bretherton2005energy,muller2012detailed,jeevanjee2013convective,coppin2015physical,muller2015favors,bretherton2015convective,wing2017convective,holloway2017observing}.
Explanations have in common that a feedback must exist, by which already moist regions grow moister, while dry regions become even drier. 
One possible mechanism is that cloudy regions lose less heat through long-wave radiation than do cloud-free regions.
In this process, subsidence would build up over the cloud-free regions, leading to low-level divergence there and suppression of cloud and precipitation.
In some observational work, however\cite{tobin2012observational}, no strong sensitivity of the radiative budget at the top of the atmosphere to self-aggregation has been found.

Here, we use equilibrium statistical physics only as an analogy, but follow a dynamical systems approach. 
Taking large-scale organization as an emergent aspect from the small scales, we test whether interactions between convective updrafts can give rise to system-scale organization. 
Cold pools, resulting from evaporative cooling under precipitating convective clouds, spread as gravity currents away from the center of the precipitation cell. 
At the scale of $\sim 10$ $km$, cold pool gust fronts have been shown to stimulate the initiation of new convective updrafts \cite{tompkins2001organizationCold,tompkins2001organizationWater,grabowski2001coupling,terai2013aircraft,langhans2015origin,feng2015mechanisms,grant2016cold,romps2016sizes,boing2016object,de2017cold,zuidema2017survey,haerter2018intensified,torricold}.
Cold pools can reach maximal radii of $r_{max}\sim 10$---$100$ $km$ over sea surfaces, and travel at an initial radial velocity of $\sim 5\;m\;s^{-1}$, which is however gradually reduced in the course of their lifetimes \cite{black1978mesoscale,zuidema2012trade,feng2015mechanisms,romps2016sizes}.
It is increasingly appreciated that correlations exist between the locations of present precipitation cells and those of subsequent ones, likely mediated through cold pool activity \cite{boing2016object,windmiller2017organization,de2017cold,torricold,haerter2018reconciling} and that precipitation can widen the distribution function of boundary layer buoyancy \cite{haerter2018intensified}.
In recent work \cite{torricold,haerter2018reconciling}, it was shown that the collision between cold pools is the dominant process of generating new convective events --- activation by isolated gust fronts without collision may be of second order. 
Interestingly, strong cold pool activity was stated to hamper the buildup of a self-aggregated state in simulations \cite{jeevanjee2013convective}, while cold pools are indeed ubiquitous in regions of precipitating convection \cite{zuidema2017survey}, in particular in cases of deep convection.

We therefore ask, how much a convective cloud system can self-organize into an aggregated state, if cold pool interactions are the relevant process leading to new convective cells?
In recent conceptual and simulation work \cite{haerter2018reconciling}, the detailed interactions between cold pools were considered. 
It was shown, by the analysis of large-eddy simulations, that it is most often collisions between several colliding cold pool gust fronts, rather than individual gust fronts, that set off new convective updrafts.
In describing the positions of new cells by a simplified geometrical model, positions were taken to be precisely defined, e.g. when growing cold pool gust fronts collided, the new cell was produced at the first point of intersection between the gust fronts. 

Here, we first relax the requirement that new cells must occur at precisely defined points. 
In other words, we take new cells to emerge {\it near} gust front collisions of existing cold pools, but allow the new cells to be displaced by a random distance of the order of the typical cold pool radius.
Such displacement has indeed been mentioned in early work \cite{tompkins2001organizationCold}, where new cells emerge somewhat randomly, but generally along the line of collision between two previous cold pool gust fronts. 

Such spatial randomness allows us to start from the assumption that cold pool interactions can be modeled by only the number density, rather than explicit positions, of cell centers in a given local environment.
The continuum model we derive enables us to map out a phase diagram of aggregation.
In a second step, we again tighten the assumptions, and return to microscopic correlations, where new cells do have a precisely-defined position relative to the ones causing them.
We contrast this refined model to the continuum model.
Our results support the key conclusion that, in both types of models, collisions between cold pools can lead to convective self-aggregation while single cold pool triggering alone cannot.
Radiative effects are only required in a domain-average sense, in order to constrain the total system energy flux.

\section*{Results}\label{sec:results}

\begin{figure*}[!]
\begin{center}
\includegraphics[height=7cm]{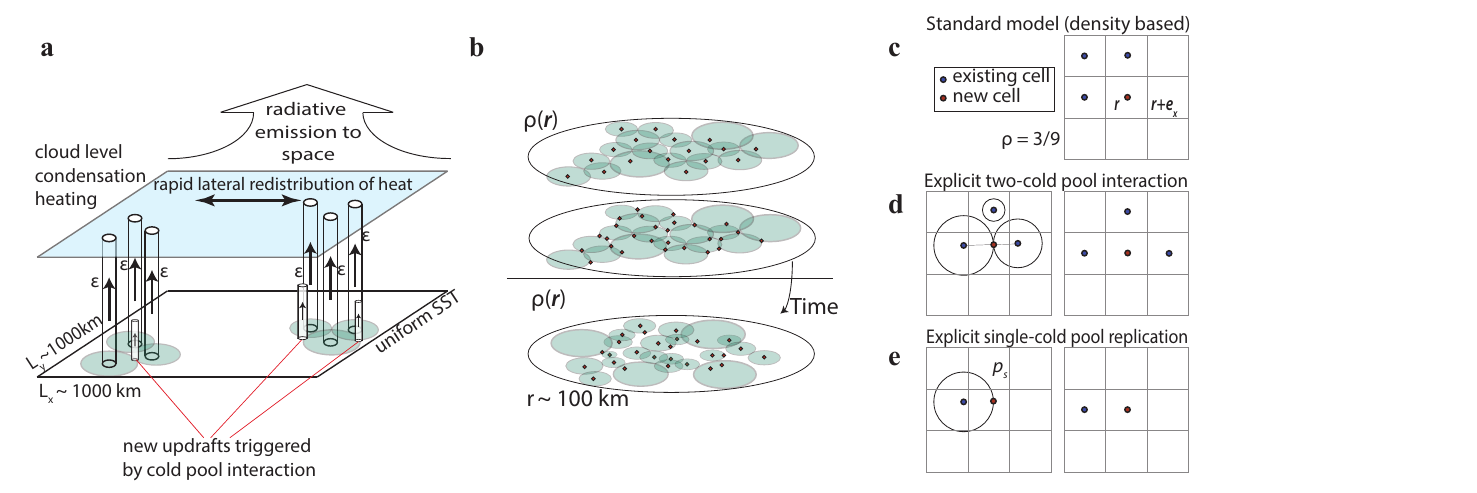}

\caption{{\bf Simple model for organization through cold pools.}
{\bf a}, Schematic of cloud system, with vertical columns showing updrafts of energy flux $\epsilon$ each, cold pools (green ellipses at the surface), new cells created by cold pools (small columns), as well as the cloud level. 
{\bf b}, Details of local cold pool interaction (lateral scale $\sim 100\;km$): schematic shows precipitation cell centers (red points) and their corresponding cold pools (green), where at a given timestep cold pools spread and form new cells near points of interference.
$\rho(\mathbf{r})$ is the number density of cells within the sub-domain shown. 
The process continues at the subsequent timestep, with cold pools again emanating from the new cell centers. 
{\bf c}---{\bf e}, Distinguishing processes:
{\bf c}, Purely density based formulation: the probability of the central lattice site becoming occupied is proportional to the density $\rho(\mathbf{r})$ of occupied sites in its surroundings. 
In the example shown, $r_{max}$ is chosen so that all eight neighboring sites contribute to the density, that is, $\rho(\mathbf{r})=1/3$, since three neighboring sites are occupied (blue).
{\bf d}, Explicit two-cold pool interaction: a new cell can be set off only if the lattice site $\mathbf{r}$ lies between any of the neighboring lattice sites. In the example shown, a new cell is possible, since blue cells lie both to the left and right of the site $\mathbf{r}$.
The circles indicate possible cold pool gust fronts. 
{\bf e}, Explicit single-cold pool replication: a new cell is possible within the neighborhood of an existing cell. 
}
\label{fig:simple_RCE}
\end{center}
\end{figure*}

\noindent
{\bf Model}\\
As our conceptual model aims to incorporate a global energy constraint, the heat entering the atmosphere through the lower boundary (e.g., a tropical sea surface) should leave at the top.
Our model considers this energy flux by assuming that heat is transported to the cloud layer through latent heating during cloud formation (Fig.~\ref{fig:simple_RCE}a).
It further assumes that heat within the cloud layer will equilibrate relatively fast with the surroundings through comparably rapid gravity waves \cite{bretherton1989gravity}. 
All heat transported to the cloud layer is eventually radiated out to space by long-wave radiative emission.
We simply mimic the global energy constraint by assigning an equal energy flux $\varepsilon$ to each convective cloud and refer to the total system energy flux as $\mathcal{E}$.

The model domain is a square of linear dimensions $L_x=L_y=L$ and an effective elementary area $a$ is required for each precipitation cell. 
$a$ necessarily includes the precipitation cell itself, but also further effects that inhibit other precipitation cells from populating this area at the same time. 
This could be downbursts or strong temperature depressions caused near the precipitation cell.
We additionally define a (generally larger) area unit, $a_{cp}=\pi\;r_{max}^2$ as the area that a cold pool gust front can cover, if its maximum radius is $r_{max}$.
As mentioned, $r_{max}\sim 10$ ---$100$ $km$, which is larger than the updraft shaft and downburst area, but substantially smaller than the domain area, that is 
\begin{equation}
 a< a_{cp}\ll L^2\;.
 \label{eq:relation_parameters}
\end{equation}

The domain total number of precipitation cells is expressed as $N$. 
Specifically, listing all horizontal positions (2D) of precipitation cells as $\mathbf{c}_i$ we can write 
\begin{equation}
 N\equiv \int_{\mathbf{r}}d{\mathbf{r}}\sum_i \delta (\mathbf{r}-\mathbf{c}_i)\;,
 \label{eq:N}
\end{equation}
where $\delta$ is the Dirac delta function and the integral is taken over the entire domain.

The domain mean number density $\overline{\rho}\equiv N/N_{max}$ with $N_{max}\equiv L^2/a$ the maximal number of cells.
$\overline{\rho}$ is hence the probability of finding a given elementary area $a$ occupied by a precipitation cell.
The total energy flux then is $\mathcal{E}=N\varepsilon$, but we will generally simply work in units of total number density $\overline{\rho}$, as all our rain cells have equal energy flux.
Analogously to Eq.~\ref{eq:N}, at any given position $\mathbf{r}$ we also define a local number density 
\begin{equation}
\rho({\mathbf{r}})\equiv a\;a_{cp}^{-1}\int_{r=0}^{r_{max}}\int_{\phi=0}^{2\pi}  d r d\phi\sum_i \delta (\mathbf{r}-\mathbf{c}_i)\;,
\label{eq:local_rho}
\end{equation}
which specifies the number density of cold pools which may affect the point $\mathbf{r}$ (Fig.~\ref{fig:simple_RCE}b,c), since the integral in Eq.~\ref{eq:local_rho} regarding $r$ is taken within the range $0<r<r_{max}$, that is, the maximal radius within which one cold pool can collide with another.
Only to avoid possible confusion, we emphasize that the symbols $\rho$ and $\overline{\rho}$ are here used as number densities, whereas, in other contexts, the symbol $\rho$ sometimes denotes the density of air within and surrounding cold pools.

To describe the dynamics of $\rho(\mathbf{r})$ we consider three dynamical processes: 
(i) spontaneous cell production; (ii) cell decay; (iii) cell interactions.

\begin{itemize}
 \item {\it spontaneous cell production} occurs for sufficient atmospheric instability and space for cells to emerge. To incorporate the former, the rate of spontaneous production is tied to the total energy flux by making the spontaneous production $\sim f_{sg}\;(1-\overline{\rho})$. Lower rates of spontaneous production hence occur when the domain mean flux is already large. $f_{sg}\geq 0$ is thereby a parameter controlling the disorder, or random seeding of cells, in the system. 
As downdrafts and cooling through rain evaporation effectively reduce the buoyancy near a given cell center after a precipitation event has occurred \cite{feng2015mechanisms,moseley2018statistical}, a limitation on the space available to new cells should be incorporated in the model.
This is accomplished by the additional factor $1-\rho(\mathbf{r})$, which ensures that each vacant area $a$ has the same probability of experiencing a spontaneous event. In total, spontaneous growth evolves as $f_{sg}(1-\overline{\rho})(1-\rho(\mathbf{r}))$.
 \item {\it cell decay} occurs for each cell present at an equal rate $f_{d}\equiv \tau_{d}^{-1}$, where $\tau_{d}>0$ is the effective duration of a precipitation event (in practice\cite{feng2015mechanisms,moseley2018statistical}, $\tau_d$ is on the order of $1$ $hour$). Local cell density $\rho(\mathbf{r})$ hence evolves as $-f_d\;\rho(\mathbf{r})$, leading to exponentially decaying cell populations if all other processes were absent. 
 \item {\it cell interaction} is due to cold pool processes. 
 For all vacant areas in the vicinity of $\mathbf{r}$, i.e. $\sim (1-\rho(\mathbf{r}))$, the cold pools emanating from all cells present within a radius $r_{max}$ around $\mathbf{r}$ can help instigate a new cell at $\mathbf{r}$. 
 We therefore introduce an effective probability $p_0\geq 0$ for cold pools in the vicinity to collide at  position $\mathbf{r}$. 
 The cell production from cold pool interaction is then modeled by $P_0(\overline{\rho})\rho(\mathbf{r})^m(1-\rho(\mathbf{r}))$, where the probability $P_0(\overline{\rho})=p_0(1-\overline{\rho})$ again warrants the total energy constraint.
 Large values of $p_0$ mean, that cold pools are more efficient at generating new cells.
 It is important to understand the meaning of the exponent $m$, which serves to distinguish cold pool processes resulting from one or multiple cold pools: 
 $m=1$ describes a process where replication is proportional to the density $\rho(\mathbf{r})$, that is, each cold pool replicates at a rate that is independent of the presence of others. 
 $m=2$ describes processes that involve collisions, that is, the replication of one cold pool depends on the presence of others in the surroundings.
  The general ability of cold pool collisions to trigger new convective cells \cite{droegemeier1985three,tompkins2001organizationCold,feng2015mechanisms,torri2019cold},
 and the explicit comparison between $m=1$ and $m=2$ suggest a crucial role of collision effects between distinct cold pool gust fronts.\cite{haerter2018reconciling}

\end{itemize}
Note that our key assumption, which will be relaxed later, is that cold pool interaction is a purely density dependent process. 
Geometrical constraints and morphologies of cell organization are hence considered higher-order corrections.
Such an assumption of a {\it density field} would be justified, when collisions between cold pools are sufficiently noisy, meaning that new cells are never produced precisely at the position of collision. 
Hence, in our model the number of cold pools contributing to the production of a new cell is relevant, but not the detailed position of the cell produced --- justifying a coarse-grained density field $\rho(\mathbf{r})$ near the point of collision. 

Piecing together the three processes above, the complete dynamical equation describing the evolution of cell density $\rho(\mathbf{r},t)$ is
\begin{eqnarray}
 \frac{d}{dt}\;\rho(\mathbf{r},t)&=&p_0(1-\overline{\rho})\rho(\mathbf{r})^m(1-\rho(\mathbf{r}))+f_{sg}(1-\overline{\rho})(1-\rho(\mathbf{r}))-f_d\;\rho(\mathbf{r})\\
 &=&(1-\overline{\rho})(1-\rho(\mathbf{r}))(p_0\;\rho(\mathbf{r})^m+f_{sg})-f_d\;\rho(\mathbf{r})\;.
 \label{eq:cell_evolution}
\end{eqnarray}
For subsequent use, we express all rates in units of $f_d$.
This is accomplished by dividing Eq.~\ref{eq:cell_evolution} through by the precipitation frequency $f_d$ (inverse duration of a precipitation event). 
This amounts to the replacements $p_0\rightarrow p_0/f_d$, $f_{sg}\rightarrow f_{sg}/f_d$, $f_d\rightarrow 1$, leaving $p_0$ and $f_{sg}$ as the remaining parameters.
Time is now measured in units of precipitation duration, and space in units of $a$ (defined above).
Furthermore, we define $q\equiv 1-\overline{\rho}$ and, from now on, drop the explicit reference to the argument $\mathbf{r}$ of $\rho$, taking the symbol $\rho$ to always refer to local cell density, as opposed to $\overline{\rho}$, which measures the system average density.
We can then more compactly define the RHS of Eq.~\ref{eq:cell_evolution} as
\begin{eqnarray}
 F_{q,p_0,f_{sg}}(\rho)&\equiv& q\;(1-\rho)(p_0\;\rho^m+f_{sg})-\rho\;.
 \label{eq:F_qpf}
\end{eqnarray}
To make the analogy to a critical phenomenon more explicit, we note that, as a polynomial in $\rho$, Eq.~\ref{eq:F_qpf} can be interpreted as the derivative of a potential, denoted $V_{q,p_0,f_{sg}}(\rho)$, where $F_{q,p_0,f_{sg}}(\rho)=-d\;V_{q,p_0,f_{sg}}(\rho)/d\;\rho$. 
For $m=2$, $V_{q,p_0,f_{sg}}(\rho)$ becomes of fourth order in $\rho$ and can exhibit two competing local mimima.\\
A few comments are appropriate:\\
\noindent
(i) the growth limitation by the factor $q=1-\overline{\rho}$ is analogous to that for logistic growth in population dynamics, where the term is interpreted as a total resource limitation; the term $(1-\rho)$ has a similar effect, but encodes local space limitations. The factor $(1-\rho)$ takes into account that two cold pools cannot be in the same location, which is reminiscent of the repelling force in a Van der Waals Gas of particles with a finite radius.
Conversely, for $m=2$ the generation of new cold pools by $\rho^m$ in Eq.~\ref{eq:F_qpf} is promoted by having two cold pools in each other's vicinity, giving a competing effect between space limitation and growth. 

\noindent
(ii) the factor $p_0\;\rho^m+f_{sg}$ is crucial in describing the local dynamics of cell growth. 
For sufficiently low density $\rho$, spontaneous cell production is dominant, while for large $\rho$, most new cells are produced by cold pool interactions; 

\noindent
(iii) for bistable steady state solutions to Eq.~\ref{eq:cell_evolution}, the exponent $m$ should be larger than unity. 
To see this, consider that, for $m=1$, $F_{q,p_0,f_{sg}}(\rho)=q(1-\rho)(p_0\rho+f_{sg})-\rho$ is quadratic in $\rho$ and therefore can only have a single stable fixed point.
More explicitly, $F_{q,p_0,f_{sg}}(1)=-1$ and $F_{q,p_0,f_{sg}}(0)=qf_{sg}\geq 0$.
Hence, depending on the choices of $p_0$ and $f_{sg}$, there is (a) either a unique stable fixed point at $\rho\leq 0$ (negative $\rho$ is even unphysical) or (b) a unique stable fixed point at a density $0<\rho<1$.
For $m=2$, which we will discuss in detail in the following, the steady state expression of Eq.~\ref{eq:cell_evolution} is of third order in $\rho$ and bistability becomes possible for certain parameter combinations. 

\noindent
{\bf Defining self-aggregation}\\
In some studies, increases in the spatial variance of cloud or liquid water has been employed as a measure of self-aggregation.
However, this measure would still allow for many disconnected cloud clusters.
Here, we want to define self-aggregation as the state where, in the limit of $t\rightarrow \infty$, complete segregation of the domain into a cloudy and a non-cloudy component takes places. 
To quantify this, we will subsequently use the number densities of convective cells within the different phases. 

\noindent
{\bf Case without spontaneous generation}\\
Consider the case of $f_{sg}=0$, a situation where new cells are exclusively produced by cold pool collisions. 
We are interested in the range of parameters, where a fully aggregated state can be stable. 
To make progress, we check for the stability of two bulk phases and the boundary between these bulk phases: 
the two bulk phases consist of the stable solutions to Eq.~\ref{eq:cell_evolution}, where the density is either high or low.
Using that $f_{sg}=0$, Eq.~\ref{eq:cell_evolution} simplifies to
\begin{eqnarray}
  \dot{\rho}&=&F_{q,p_0}(\rho)\;,\\
  F_{q,p_0}(\rho)&\equiv&-\rho\;q\;p_0\;(\rho^2-\rho+1/(q\;p_0))\;.
 \label{eq:cell_evolution_simple}
\end{eqnarray}
Looking for local solutions in dependence on $q$, apart from the trivial solution $\rho_1=0$, we obtain
\begin{equation}
 \rho_{2,3}=\frac{1}{2}\mp\left(\frac{1}{4}-\frac{1}{p_0q}\right)^\frac{1}{2}\;.
 \label{eq:rho_23}
\end{equation}
Eq.~\ref{eq:rho_23} offers physically meaningful solutions for $p_0q\geq 4$.
Both $\rho_1$ and $\rho_3$ are stable fixed points, which we check by verifying that $\partial F_{q,p_0}/\partial \rho|_{\rho=\rho_{1,3}}<0$.

\begin{figure*}[t]
\begin{center}
\includegraphics[height=6cm]{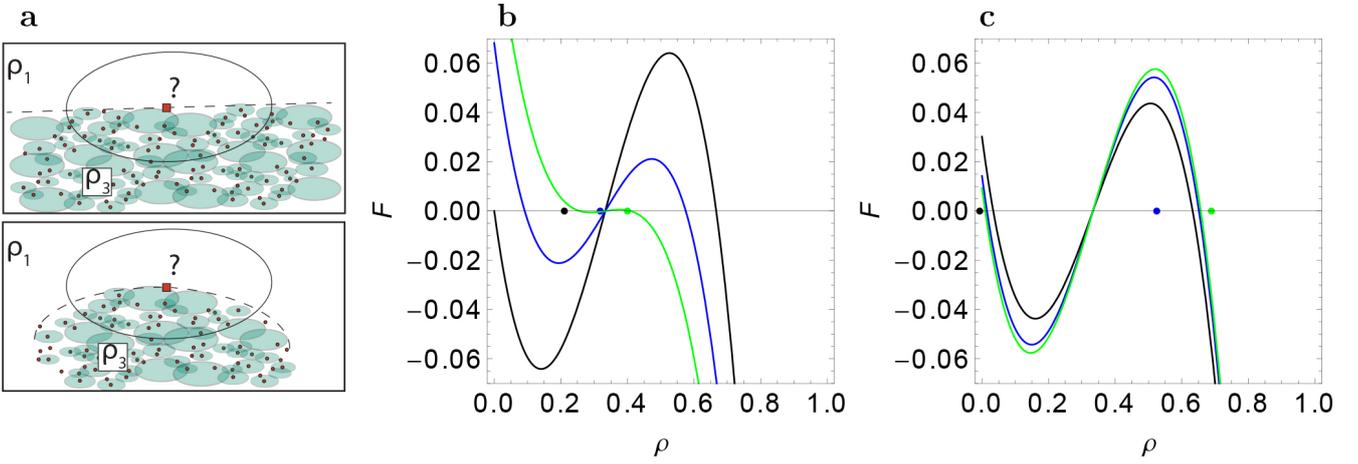}

\caption{{\bf Fixed points for different models.}
{\bf a}, Schematic exemplifying the effect of an infinite boundary (approximately straight line boundary), vs a finite boundary, which then shows a pronounced curvature affecting the replication probability at the boundary.
The red square indicates a point at the boundary, which uses the density computed within the circled area, to determine a possible update.
{\bf b}, Model with quadratic dependence on $\rho$, i. e. $m=2$ in Eq.~\ref{eq:cell_evolution}, where $q=9/2(9f_{sg}+p_0)$ is chosen to ensure that $\rho_1+\rho_3=2\rho_2$, i.e., that an aggregated state is possible.
Black, blue and green curves correspond to increasing values of $f_{sg}=\{0,0.1,0.2\}$, respectively, with fixed $p_0=5.7$. 
Small round symbols in corresponding colors along the horizontal axis indicate the corresponding values of $\overline{\rho}=1-q$.
Note that the value of $\overline{\rho}$ increases with $f_{sg}$ and eventually departs from the allowed range $\rho_1<\overline{\rho}<\rho_3$.
{\bf c}, Similar to (a) but for $f_{sg}=.03$ held fixed and $p_0=\{4.2,9.2,14.2\}$ for colors black, blue and green, respectively.
Note that in this example only the intermediate value of $p_0$ yields a self-aggregated state.
}
\label{fig:fixed_points}
\end{center}
\end{figure*}

So far, we have treated $q\equiv 1-\overline{\rho}$ as a parameter. 
However, $q$ should be obtained in such a way that the partitioning of the domain into regions of high and low density ($\rho_3$ and $\rho_1$, respectively) is stable.
One therefore has to consider the interface between these two regions of densities $\rho_1$ and $\rho_3$, respectively, and demand that the density at the interface be constant (Fig.~\ref{fig:fixed_points}a).
As an approximation, consider the interface to locally be a straight line boundary, that is, we neglect its curvature and consider that the interface is smooth and sharp, so that for one half plane $\rho=\rho_1=0$, for the other, $\rho=\rho_3$, as given by Eq.~\ref{eq:rho_23}.
This straight line approximation is valid, as long as $r_{max}$ is far smaller than the radius of the aggregated sub-domain.
For finite-size systems the curvature can generally not be neglected and should be treated as a correction.
We describe the effective density for any point at the interface as the average $\rho_{ave}\equiv (\rho_1+\rho_3)/2=\rho_3/2$, and further that $\rho_{ave}$ be constant, i.e.
\begin{equation}
F_{q,p_0}(\rho_{ave})=0\;, 
\end{equation}
or, equivalently,
\begin{equation}
 \rho_2=\rho_3/2\;.
\end{equation}
Using this together with Eq.~\ref{eq:rho_23} yields
\begin{equation}
 q=1-\overline{\rho}=\frac{9}{2}p_0^{-1}\;,
 \label{eq:q_vs_p}
\end{equation}
and $\rho_1=0$, $\rho_2=1/3$, $\rho_3=2/3$.
Hence, if the aggregated state exists, the total cloud-free area $q$ is inversely proportional to the interaction factor $p_0$ and the density within the cloudy bulk area is independent of $q$ and equal to $2/3$.
But for which $p_0$ can aggregation be expected?
Clearly, $\overline{\rho}=1-q$, the average density, must lie between the densities of the two phases, hence, 
\begin{equation}
\rho_1< 1-q< \rho_3\;,
\label{eq:condition_q}
\end{equation}
yielding
\begin{equation}
 \frac{9}{2}\equiv p_{0,min}<p_0<p_{0,max}\equiv \frac{27}{2}\;.
\label{eq:pmax}
 \end{equation}

\begin{figure*}[t]
\begin{center}
\includegraphics[height=7cm]{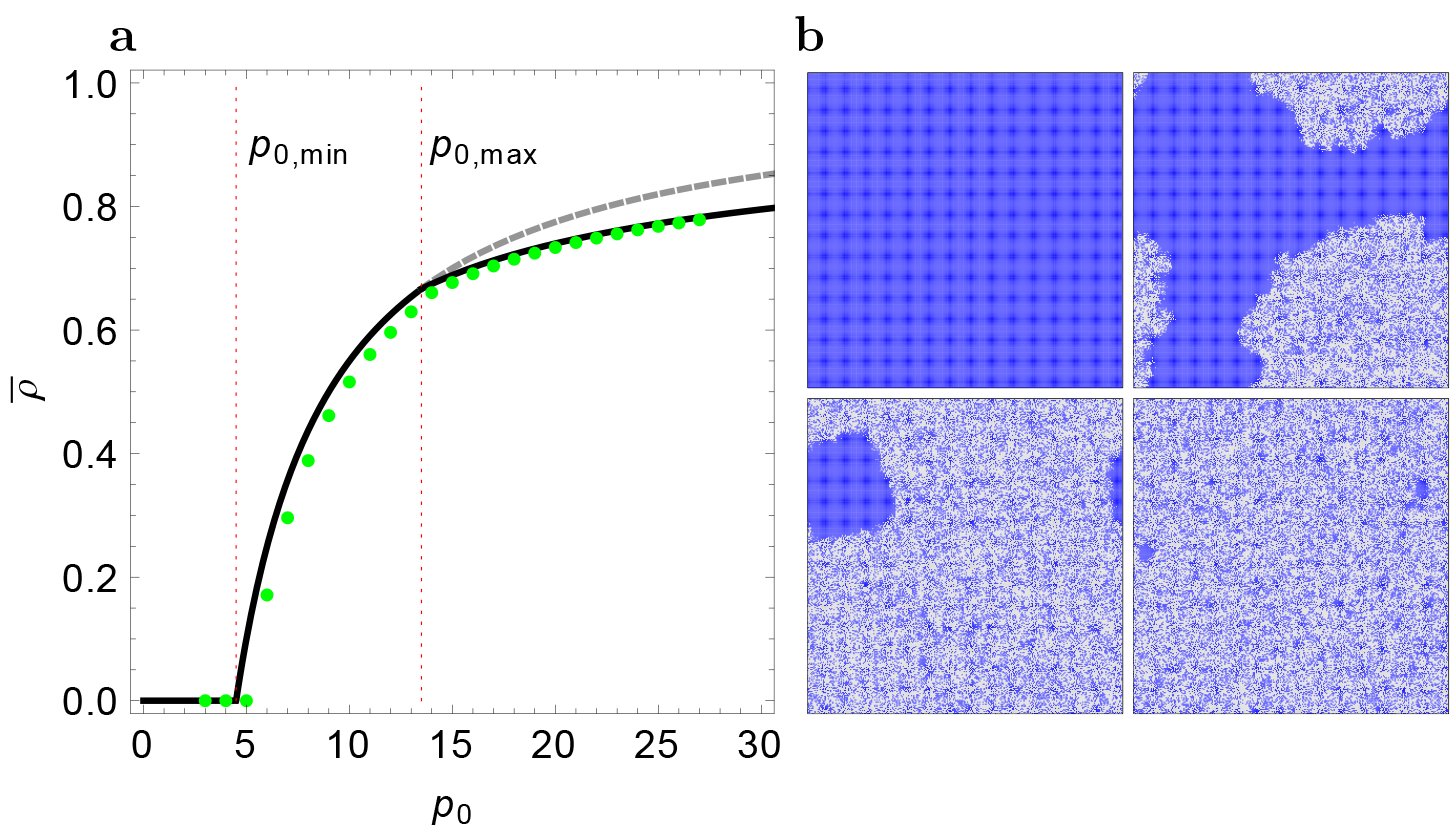}
\caption{{\bf Cloud fraction vs. interaction.}
{\bf a}, Cloud fraction $\overline{\rho}$ as a function of $p_0$ in the case without spontaneous generation ($f_{sg}=0$). 
Theoretical and simulation result shown as black line (green points). 
Red dotted lines indicate $p_{0,min}$ and $p_{0,max}$, respectively, that is, the limits between which segregation can take place.
Gray dashed line indicates the function $\overline{\rho}=1-9/2p_0$ (valid in the segregated regime), to indicate the change of slope at $p_{0,max}$.
The slight discrepancy between the theoretical and simulation results is likely due to the assumption of smooth and straight interfaces between the cloudy and cloud-free areas, as well as noise caused by finite size effects and lattice discretization.
Simulations were carried out on a lattice of $300\times 300$ sites with periodic boundary conditions using a Gillespie algorithm for all rates involved, several thousand system updates per parameter value were simulated before computing the average $\overline{\rho}$.
{\bf b}, Plots indicating the spatial pattern of different steady states, cloudy (gray) and cloud-free regions (blue), for $p_0=\{4,7,13,14\}$, hence cloud-free, aggregated with considerable cloud-free areas, aggregated but mostly cloudy, and fully cloudy.
}
\label{fig:aggregation_p0}
\end{center}
\end{figure*}

Hence, the segregated phase is limited to intermediate values of $p_0$.
For $p_0\leq p_{0,min}$, the domain is entirely cloud-free, that is, $\overline{\rho}=1-q=0$.
For $p_0\geq p_{0,max}$, no long-lived cloud-free region will exist. 
However, transient density fluctuations are possible. 
Our solution predicts that, for $p_0\geq p_{0,max}$, $\overline{\rho}$ will follow the uniform bulk solution (inserting $\rho=\overline{\rho}=1-q$ and $f_{sg}=0$ in Eq.~\ref{eq:F_qpf}) to
\begin{equation}
 F(\overline{\rho})=p_0(1-\overline{\rho})^2\overline{\rho}^2-\overline{\rho}=0\;,
 \label{eq:F_bulk}
\end{equation}
and it is easy to check that for $p_0\geq p_{0,max}$ there is only one stable, plausible, solution for $\overline{\rho}$.
Eq.~\ref{eq:F_bulk} gives $\overline{\rho}(p_{0,max})=2/3$, which matches that obtained from Eq.~\ref{eq:q_vs_p} for the segregated solution, hence $\overline{\rho}(p_0)$ is continuous at $p_{0,max}$.
We further check the slopes 
\begin{eqnarray}
s_< &\equiv &\lim_{p_0\rightarrow p_{0,max} \atop p_0<p_{0,max}}\frac{\partial \overline{\rho}}{\partial p_0}=\frac{2}{81}\;\\
s_> &\equiv &\lim_{p_0\rightarrow p_{0,max} \atop p_0>p_{0,max}}\frac{\partial \overline{\rho}}{\partial p_0}=\frac{4}{243}\;,\\
\label{eq:limits_s}
\end{eqnarray}
hence, $s_< >s_>$, signaling a continuous phase transition of $\overline{\rho}(p_0)$ at $p_{0,max}$.
We check this result by explicit simulations (Fig.~\ref{fig:aggregation_p0}).

\begin{figure*}[t]
\begin{center}
\includegraphics[height=6cm]{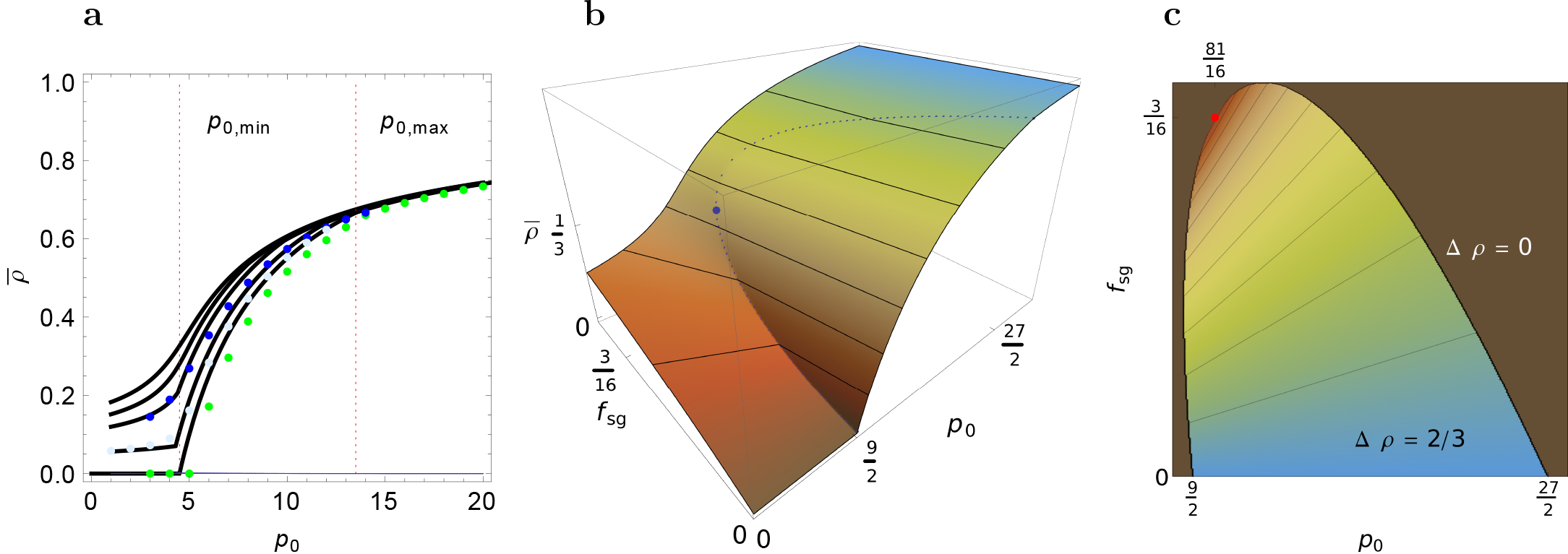}
\caption{{\bf Phase diagram for a system with spontaneous generation.}
{\bf a}, Cloud cover versus interaction strength $p_0$. 
Similar to Fig.~\ref{fig:aggregation_p0} but now allowing for finite spontaneous generation $f_{sg}\geq 0$.
Black lines show theoretical results, points of varying colors indicate simulation results on a domain of linear dimension $L=200$, with simulations carried out as in Fig.~\ref{fig:aggregation_p0}). 
Curves from bottom to top correspond to $f_{sg}=\{0,.06,.14,3/16,.24\}$, that is, for the first three we expect a change of slope as function of $p_0$, for the remaining two we expect smooth behavior.
For orientation, the dotted lines indicate the values $p_{0,min}$ and $p_{0,max}$ for the case of $f_{sg}=0$.
{\bf b}, Surface plot showing domain average cloud fraction $\overline{\rho}$ versus interaction strength $p_0$ and spontaneous generation rate $f_{sg}$. 
Color shading from orange to blue indicates increasing values of $\overline{\rho}$. 
Thin black lines are contour lines for several values of $\overline{\rho}$.
Dotted line indicates the boundary between non-aggregated (continuum phase) and aggregated convection (clustered phase).
Note the abrupt change of slope at the dotted line (most easily visible for $p_0\approx 9/2$.
Note also the ``triple point'' at $(81/16,3/16,1/3)$, marked as a larger blue point, which is characteristic, as it is located at a smooth transition between the continuum and clustered phase. 
{\bf c}, Heatmap of the contrast $\Delta \rho\equiv \rho_3-\rho_1$, between the densities within and outside the aggregated subdomain.
Fig.~S1 shows examples of spatial patterns.
}
\label{fig:3d_case}
\end{center}
\end{figure*}

\noindent
{\bf Case with spontaneous generation}\\
We now allow spontaneous generation ($f_{sg}>0$) and work with the full Eq.~\ref{eq:cell_evolution}, namely
\begin{eqnarray}
 \dot{\rho}&=& F_{q,p_0,f_{sg}}(\rho)\;,\\
F_{q,p_0,f_{sg}}(\rho)&\equiv &-qp_0\rho^3+qp_0\rho^2-\rho (1+qf_{sg})+q\;f_{sg}\;.
\label{eq:full_equation}
 \end{eqnarray}

One obvious effect of $f_{sg}>0$ is to increase the value of the lower stable fixed point, that is, $\rho_1\approx qf_{sg}$ to linear order, making $\rho_1$ positive.
As a third order polynomial, the zeros of Eq.~\ref{eq:full_equation} are more complicated algebraic expressions ({\it Details:} Supplement). 
However, to test for stability of the aggregated state we are again mainly interested in fixed points, where 
\begin{equation}
 \rho_2=(\rho_1+\rho_3)/2\;
 \label{eq:rho_2_conditions}
\end{equation} 
is fulfilled by appropriate adjustment of $q$. 
Luckily, the condition on $q$ becomes very simple, namely
\begin{equation}
 q=\frac{9}{2(9f_{sg}+p_0)}\;,
 \label{eq:q_general}
\end{equation}
which yields Eq.~\ref{eq:q_vs_p} for $f_{sg}=0$, as it should.
The resulting zeros are also simple, namely 
\begin{eqnarray}
 \rho_{1/3}&=& \frac{1}{3}\left( 1\mp \left(1-\frac{27f_{sg}}{p_0}\right)^{1/2}\right)\;,\label{eq:rho_13_gen}\\
 \rho_2&=&1/3\;.\nonumber
\end{eqnarray}

We can quantify the {\it contrast} $\Delta\rho(p_0,f_{sg})$ between the densities in the aggregated and cloud-sparse regime, as
\begin{equation}
\Delta\rho(p_0,f_{sg})\equiv\rho_3-\rho_1=\frac{2}{3}\left(1-\frac{27f_{sg}}{p_0}\right)^{1/2}\;,
\label{eq:contrast}
\end{equation}
which is sharp for large interaction strength $p_0$ and small $f_{sg}$.

However, we have not yet established where this contrast is applicable, as we need the boundaries of the aggregated regime.
To obtain this boundary, we again use the condition on $q$ (Eq.~\ref{eq:condition_q}), which together with Eqs~\ref{eq:q_general} and \ref{eq:rho_13_gen} yields the upper and lower lines of transitions $p_0(f_{sg})$ (Eqs S4 and S5).
While these are more elaborate algebraic expressions, they are nonetheless closed and allow us to study the type of transition occurring at the boundaries of the aggregated regime.

Evaluating the expressions for the boundary (Eqs~S4,S5) together with Eq.~\ref{eq:q_general}, we first compare the previous results for $f_{sg}=0$ (Fig.~\ref{fig:aggregation_p0}) with those of positive $f_{sg}$ (Fig.~\ref{fig:3d_case}a).
The primary effect of increased $f_{sg}$ is to increase total density $\overline{\rho}$.
However, for small values of $f_{sg}$, the slope still changes abruptly when increasing $p_0$ beyond a threshold. 
This continuous phase transition signals the entry to the aggregated state (Eqs~S4 and S5). 
For sufficiently large $f_{sg}$ a change of slope as function of $p_0$ is no longer observed.

Why does a change of slope occur?
Within the aggregated phase, an additional degree of freedom is activated, namely one where cells lump together to ``aid'' one another, thus leading to more favorable conditions for replication.
At larger values of $f_{sg}$ the discontinuity vanishes --- abrupt variations in state are no longer possible. 
This is explained by $f_{sg}$ diluting free space so strongly that new aggregates will incessantly form all over the domain (compare: Fig.~\ref{fig:fixed_points}b).

The figure also shows simulation results for the different parameter combinations. 
The contrast between theory and simulation is strongest for $f_{sg}=0$, but also the cases with spontaneous generation display systematic discrepancies.
These are partially due to finite system-size effects, whereby at the transition to the aggregated state very small clusters will initially form. 
These clusters suffer most from the curvature effect (Fig.~\ref{fig:fixed_points}a), hence, small clusters will enter a positive feedback loop of decay, where they become smaller and increase the detrimental curvature effect.

Also for larger $p_0$, at the other end of the aggregated regime, a discontinuity occurs, which is however less noticeable. 
At this stage, the domain is so densely filled that interaction takes place all over, and clustering is no longer possible. 
The slope in fact decreases when crossing the threshold (Eq.~S5), a bunched-up state could maintain higher activity, but it becomes statistically impossible to maintain such a segregated state.
We summarize the findings for $\overline{\rho}(p_0,f_{sg})$ (Fig.~\ref{fig:3d_case}b) and $c(p_0,f_{sg})$ (Eq.~\ref{eq:contrast}, Fig.~\ref{fig:3d_case}c).

In summary, edges of the segregated phase within the two-dimensional phase diagram are generally characterized by continuous transitions in $\overline{\rho}$.
However, for a single combination of $p_0$ and $f_{sg}$ this change of slope should disappear, namely, when $\rho_1=\rho_2=\rho_3=\overline{\rho}$ ({\it compare}: Fig.~\ref{fig:fixed_points}). 
This point is easy to obtain when using Eq.~\ref{eq:q_general} and exploiting the equality of densities in Eq.~\ref{eq:rho_13_gen}.
We thus obtain the ``triple point'' as
\begin{eqnarray}
 f_{sg}^*&=&\frac{3}{16}\;,\\
 p_0^*&=&\frac{81}{16}\;,\\
 q^*&=&2/3=1-\overline{\rho}\;,
\end{eqnarray}
that is, a point that is characterized by a smooth transition between the continuum and the segregated phases (marked in Fig.~\ref{fig:3d_case}b,c).
At this point, the contrast (Eq.~\ref{eq:contrast} and Fig.~\ref{fig:3d_case}c) vanishes.

\noindent
{\bf Including explicit spatial interaction.}\\
As mentioned, our model has assumed some spatial displacement upon collision of cold pools, leading to a formulation where only the number density, not the explicit position of cells in the surroundings of $\mathbf{r}$ was relevant in determining interaction effects between cold pools (Fig.~\ref{fig:simple_RCE}c).
The details of how often and at which precise location cold pools trigger new convective cells is still an open research question.
Literature however states that cold pools reach typical maximal radii \cite{black1978mesoscale,zuidema2012trade,feng2015mechanisms,romps2016sizes} and that cold pool interactions often leave behind a new cell near the location of collision \cite{holle1980tornado,droegemeier1985three,tompkins2001organizationCold,torricold,feng2015mechanisms,haerter2018reconciling,torri2019cold} --- hence, requiring the new cell to lie somewhat in between the centers of the cold pools causing it.


To qualitatively capture such geometric spatial effects, we now consider a refined process on a lattice, where each lattice site represents the area $a^*$, required by a precipitation cell and its cold pool. 
In practice, the value of $a^*$ will be a result of a self-organization process, which can be understood as follows:
neighboring precipitation cells can instigate new cells, when their respective cold pools have each traveled more than a minimal distance, defined by the radius of the effective precipitation cell area $a$, but less than the maximal distance $r_{max}$ a cold pool can travel. 
Typical distances, which define $a^*$, will lie at intermediate values, and their calculation is a non-trivial statistical mechanics problem, left to a future study.
In the current qualitative discussion we take $a^*$ to be a given system parameter.

Given that the site at $\mathbf{r}$ is not occupied, and two neighboring, and spatially opposite sites are occupied by existing cells, e.g., cells at $\mathbf{r}-\mathbf{e}_x$ and $\mathbf{r}+\mathbf{e}_x$ (Fig.~\ref{fig:simple_RCE}d), a new cell can be produced at $\mathbf{r}$ by collision of the cells.
Since we only consider neighboring sites for the interaction, we are implicitly assuming that $r_{max}$ is on the order of a lattice site.
Note that this type of interaction, where the resulting cell is always produced {\it between} existing cells, will, for geometrical reasons, be unable to allow boundaries between cloudy and cloud-free areas to grow in favor of the cloudy areas (compare: Fig.~\ref{fig:fixed_points}a). 
In fact, the boundary will gradually retreat and cloud-free ``cavities'' will tend to increase in size.
A single cavity would then grow larger until it takes up the entire domain --- no sustainable aggregation could take place.
Such a run-away effect will not be physically plausible, since the convective instability would inevitably increase by surface heating and cloud-level radiative cooling. 
Eventually, other processes, e.g. the spontaneous ones discussed throughout the text, or single-cold pool processes (Fig.~\ref{fig:simple_RCE}e) would become active.

We therefore now consider an alternate model, where, in addition to the two-cold pool process described, also single cold pool processes are incorporated (Fig.~\ref{fig:simple_RCE}d,e), and the probability for a given vacant site to be ``infected'' by any occupied neighbor is $p_s$.
We remind the reader that single cold-pool processes alone were ruled out on theoretical grounds to give aggregation (discussion following Eq.~\ref{eq:F_qpf}).
We here test this conclusion numerically (Fig.~S2a, gray curve).
We again initialize our domain with a random seeding of cells and allow the dynamics to evolve.
Indeed, for $p_s\gtrsim .1$, $p_0=0$, the spreading from any occupied site to its eight nearest neighbors starts to outweigh decay, however, the spatial dynamics yields a random pattern without aggregated clustering (not shown).

We now set $p_0=10$, allowing for two-cold pool interaction.
Similar to simulations of convective self-aggregation \cite{wing2017convective}, the initial effect is that small sub-regions become cleared from precipitation cells. 
Some of these cleared regions eventually expand and merge with other cleared region. 
Finally, only a single cloudy patch remains, which maintains its area indefinitely.

Exploring the parameter $p_s$ systematically, we find that the domain is entirely cloud-free for small values, and a continuous transition to an aggregated regime again exists, similar to the case of $f_{sg}=0$ for the previous model (Fig.~\ref{fig:aggregation_p0}).
With a further increase in $p_s$, the area of the aggregated subregion further increases.
Finally, for very large $p_s$, aggregation is no longer possible and a featureless cloudy regime is found.
For this high-density limit, we could however not detect any phase transition for the order parameter $\overline{\rho}$.
Rather, at large values of $p_s$ increasingly diffusive dynamics is found which smoothly leads to a random pattern in space.
In summary, also with a stricter condition on the location of cells resulting from the collisions, a phase transition to an aggregated regime is found, when two-cold pool interactions play a sufficient role in generating the new cells.

\section*{Discussion and Conclusion}\label{sec:discussion}
Self-aggregation in convection has drawn substantial interest due to its potential applicability to large-scale structures in tropical and subtropical cloud organization, most notably the Madden-Julian Oscillation\cite{zhang2005madden} in the Indian and Pacific Ocean, as well as the buildup of hurricanes.
Cold pool processes are now considered a crucial component in the interaction between convective clouds. 
However, their exact relation to self-aggregation has been unclear.
This may, in part, be due to the large mix of effects, all of which can contribute to the self-organization of the convective cloud field --- most prominently the effect of stronger radiative cooling of cloud-free air masses, which has been implicated in the stabilization of a larger-scale circulation pattern to reinforce aggregation \cite{wing2017convective}.

With so many effects contributing, the computational demand to map out the entire phase diagram of actual aggregation in fluid-dynamics models, is currently prohibitively large. 
Likely contributors, which would need to be explored, are the value of the sea surface temperature, the ventilation coefficient which describes rain evaporation leading to cold pool formation, model grid resolution, which influences cold pool spreading and the sharpness of gust front boundaries, the radiation and cloud microphysics schemes, as well as domain size and geometry.

This study explored the possible implications of small-scale interactions between clouds. 
These interactions are active at the spatial scale of cold pool radii, which are typically on the order of tens of kilometers.
The model imposed a large-scale energy constraint, by which the propensity of forming new precipitation events is weakened when the large-scale energy budget is used up.
Our finding is that, when perturbations are weak, in other words, when the main cause of new convective events is the interaction between previous events, mediated through their cold pools, aggregation is likely to occur. 
For larger perturbations a continuum state would be reached, where the domain shows a rather featureless mix of more cloudy and less cloudy ``patchiness''.

Notably, our cold pools have been constrained by a maximal radius $r_{max}$, which sets a scale for cold pool interaction. 
Increasing $r_{max}$ would correspond to cold pools that can travel larger distances, before their momentum decays. 
In that case, interactions would become very strong due to the larger value of $\rho(\mathbf{r})$ (Eq.~\ref{eq:local_rho}), and a segregation into an aggregated and a cloudfree phase would no longer occur.
In our model (Eq.~\ref{eq:cell_evolution}), the effect of $r_{max}$ is captured in the parameter $p_0$, and increases in $r_{max}$ would translate to larger $p_0$. 
Hence, for sufficiently large $r_{max}$, $p_0>p_{0,max}$, and the aggregation regime in Fig.~\ref{fig:3d_case} would be left.
These considerations are in line with previous findings from simulations \cite{jeevanjee2013convective}, where self-aggregation was found to be hampered by increased cold pool strength.

In conclusion, we have here introduced a continuum model for convective cell spatial number density as well as a discrete model with explicit spatial interaction, to qualitatively mimic the effect of cold pool interaction within an energy flux constrained framework.
When only single cold pools are allowed to set off new convective updrafts, self-aggregation did not occur.
Single cold pool processes display a diffusion-like dynamics, which does not constitute a basis for bistability.
Conversely, independent of the model, the interaction between multiple cold pools can indeed give rise to sustained aggregation effects, which self-organize from an initially random cloud distribution. 


\section*{Acknowledgments}
JOH thanks S. J. Boeing, S. B. Nissen and K. Sneppen as well as the two anonymous reviewers for useful comments.
JOH gratefully acknowledge funding by a grant (13168) from the VILLUM Foundation.
This project has received funding from the European Research Council (ERC) under the European Union's Horizon 2020 research and innovation program (grant agreement no. 771859). 
No new data were used in producing this manuscript.

\clearpage
\pagebreak

 \newcommand{\noop}[1]{}

\renewcommand{\thesection}{S\arabic{section}}

\renewcommand{\thefigure}{S\arabic{figure}}
\setcounter{figure}{0}

\renewcommand{\thetable}{S\arabic{table}}
\setcounter{table}{0}

\renewcommand{\theequation}{S\arabic{equation}}
\setcounter{equation}{0}

\pagebreak
\clearpage

\end{document}